# Measuring the Effectiveness of Learning Resources Via Student Interaction with Online Learning Modules

Zhongzhou Chen and Matthew Guthrie

*Department of Physics, University of Central Florida, Orlando, 32816*

Abstract

We present a new method for measuring the effectiveness of online learning resources, through the analysis of time-stamped log data of students' interaction with a sequence of online learning modules created based on the concept of mastery learning. Each module was designed to assess students' mastery of one topic before and after interacting with the learning resources in the module. In addition, analysis of log data provides information on students' test-taking effort when completing the assessment, and learning effort when interacting with the learning resources. Combining these three measurements provides accurate information on the quality of each online learning module, as well as detailed suggestions for future improvements. The results from data collected from all 10 modules are presented in a sequence of sunburst charts, an intuitive visual representation designed to allow the average instructor to quickly grasp the key outcomes of the data analysis, and identify less effective modules for future improvements. Online learning modules can be implemented much more frequently than either clinical experiments or classroom assessments, and can provide more interpretable data than most existing online courses, making it a valuable tool for quickly assessing the effectiveness of online learning resources.

## 1 Introduction

As online learning becomes an increasingly important component of students' learning experiences in STEM courses, instructors today are more likely than ever to either create or incorporate various types of online learning resources as part of their instruction. At the same time, there is a rapid increase in both the number and the variety of openly accessible learning resources online. A simple search for "Newton's laws of motion" on either Google or YouTube will return thousands of relevant instructional websites, e-texts, videos and simulations.

Amid this boom of online learning resources, many instructors are facing a key question on a daily basis: how do we know which learning resources will be most effective in helping students learn a certain topic? More specifically, when a piece or a set of learning resource(s) are assigned to a given student population, what fraction of that population can improve their understanding or skill from interacting with these resources, and by how much? This paper introduces a new method to quickly provide detailed information on the effectiveness of online learning resources by analyzing and visualizing data from student interaction with an online mastery learning system.

Traditionally, the effectiveness of instructional methods and learning resources are most often measured through either clinical experiments or classroom assessments. In a typical clinical experiment, students are recruited as experiment subjects to complete certain learning activities in a clinical setting at one or more pre-determined location(s) and time(s). Subjects are often selected based on their performance on a pre-test, and their learning outcomes are assessed either by a post-test, or from data collected during the activities (for example, see [1–5]).

Clinical experiments can provide accurate measurements of students' behavior during interaction with learning resources, and learning outcomes after the interaction. However, conducting clinical experiments with adequate sample size requires significant time and resources, neither of which are usually available to the average instructor. Moreover, students may be more willing to engage with the instructional materials in a clinical setting than in an authentic learning setting. In some cases, learning gains observed in a clinical setting may be significantly reduced or even disappear when the same instructional material is assigned as homework [4,6].

In contrast, it is much easier for the average instructor to conduct classroom assessments as part of the regular instruction. One of the most common forms of classroom assessment for measuring the effectiveness of instruction is to administer not-for-credit pre and post-tests at the beginning and the end of a period of instruction [7–10], usually a semester, using a standard instrument such as the Force Concept Inventory [11]. Student learning gains can be measured from the difference in scores on the two tests [7,12,13].

However, the frequency at which pre/post-tests can be conducted in most cases is limited to once or twice per semester, which limits its ability to provide information on the effectiveness of individual learning resources. In addition, pre/post-tests do not provide information on the extent to which students engaged with instructional resources, which makes it harder to relate the observed change in score to the quality of specific instructional resources. Finally, variations in students' test-taking effort on not-for-credit surveys can impact the reliability of pre/post testing outcomes. Earlier studies have shown that some students simply make random guesses on some or even most of the questions on not-for-credit surveys [14–16].

With the development of online learning technology, a promising method for quickly collecting large amounts of data on student learning behavior at scale is to analyze the time-stamped log-data of student interaction with online learning systems [17–23]. One prominent advantage of this method is that time-stamped log-data can provide information on students' engagement with learning resources [24–27] and test-taking effort during problem solving [14,15,28], both of which can vary significantly between students especially in an online learning environment.

However, measuring the effectiveness of instructional resources using data from online courses also faces two challenges. First, learning outcomes are still assessed relatively infrequently in most online courses, which largely inherited their instructional design from their off-line, face-to-face predecessors [22,29,30]. Many online courses contain only one or two exams throughout the semester, while some others implement weekly quizzes. Second, very few online courses contain any form of pre-test that measures students' incoming knowledge before learning [31,32]. Both of those factors make it difficult to attribute students' assessment performance to their engagement with any particular instructional resource in the course.

Several recent studies have implemented a new type of online instructional design for the teaching of college level physics, based on the concept of "mastery learning" initially proposed by Bloom and Keller [3,33–36]. In an online mastery learning setting, students spend varying amounts of time and effort to reach mastery on a given concept or skill through interaction with instructional materials, before proceeding to the next concept in sequence. In this paper, we make the case that in addition to being an effective instructional method, the online mastery learning design can also be uniquely suitable for providing detailed and accurate information on the effectiveness of instructional resources.

As is described in detail in section 2, the current study implements the online mastery learning design through a sequence of 10 online learning modules (OLMs), with each module taking about 20-30 minutes to complete. As illustrated in Figure 1, upon opening a new module, students are required to make at least one attempt at solving the assessment problems in the module before being able to access the instructional materials contained in the module. After studying the instructional materials, students and can make additional attempts on similar assessment problems. Each student's attempts before and after interacting with the instructional material effectively serve as pre and post-tests, measuring the change in each student's understanding of a single concept or the ability to solve one type of physics problems.

The effectiveness of instructional materials contained in each module can be evaluated by analyzing data collected from the OLMs. This method combines multiple advantages of the three existing methods mentioned above, and overcomes some of their major shortcomings:

1. Each student's problem-solving ability is assessed at a much higher frequency compared to either conventional pre/post testing or online courses. Moreover, the assessments are closely coupled with students' learning processes, which increases the interpretability of the data collected.
2. Information on each student's test-taking effort during assessment attempts, as well as their level of engagement with the instructional materials, can be obtained from analyzing the time-stamped log data from the OLM system. Including both types of information into the analysis will improve the accuracy of the measurement.
3. Assigned as part of students' regular online homework, the OLMs are significantly easier to implement than clinical experiments, and the data collected reflects student behavior in an authentic learning setting.

In an earlier pilot study involving four OLMs provided to students as an optional exam preparation resource, we showed that data collected from the OLMs have higher interpretability, and conveys richer information on students' learning processes than data collected from traditional online courses or homework systems [25]. The current study expands the OLM sequence to 10 modules, which were assigned to 235 students in a calculus based introductory mechanics course as required online homework.

In sections 2 and 3, we will describe in detail the design and implementation of the OLMs and module sequence, as well as the procedures for data collection and cleaning. In section 4, the effectiveness of instructional resources in each of the 10 OLMs is evaluated in three steps: First, we will apply data analysis techniques to extract from the time-stamped log data students' assessment outcomes before and after learning (section 4.1), their test-taking effort during each assessment attempt (section 4.2), and their levels of engagement with the learning resources between assessment attempts (section 4.3). Second, the three types of information on each module are synthesized and presented in an intuitive visual representation: a sequence of 10 sunburst charts [37] (section 4.4). The sunburst charts highlight significant patterns in the data and present them in a visually salient format that is easy to understand by the average instructor. Finally, to assist the interpretation of the sunburst charts, we will carefully investigate how learning from instructional materials affects assessment outcomes, or more precisely, the extent to which different levels of engagement with the instructional materials in each module are correlated with different performance and behavior on the assessment attempts before and after learning (section 4.5).

In section 5, we will present our interpretation of the patterns observed from the sunburst charts and results from the correlation analysis, to gain insight into the effectiveness of each of the 10 modules. We will also discuss the advantages of the OLMs over other conventional methods, and point out the shortcomings of the current work that can be addressed in future studies.

## 2 Design of an Online Learning Module (OLM) Sequence

### 2.1 General structure of OLM and OLM Sequence

In an online mastery learning setting, students learn a new topic by completing a sequence of OLMs, each focusing on developing the understanding of one concept or practicing solving one type of problem. In the current study, the topic of mechanical energy was covered by an OLM sequence consisting of 10 modules, starting with the definition of kinetic energy and ending with problems involving changes in both kinetic and potential energy in the presence of non-conservative work (see Appendix I). In this regard, the OLM sequence also bares some similarity to deliberate practice [38,39].

Each OLM consisted of an instructional component (IC) and an assessment component (AC) (Figure 1). Upon accessing a new OLM, a student was first directed to the AC and required to make at least one attempt at the AC, which contained a short 1 or 2 problem quiz. The first page of the AC was entitled "Learning Goals" and contained a brief, one-sentence description of the concept or problem type contained in the module, and a button to start the first assessment attempt, as illustrated in Figure 2A. On this page, students were informed that if they could pass the AC by correctly solving all the problems, they would be able to directly proceed to the next module in the sequence. If not, they would be able access the IC in order to learn how to solve the problems after their first attempt, and return at any time to make additional attempts at passing the AC.

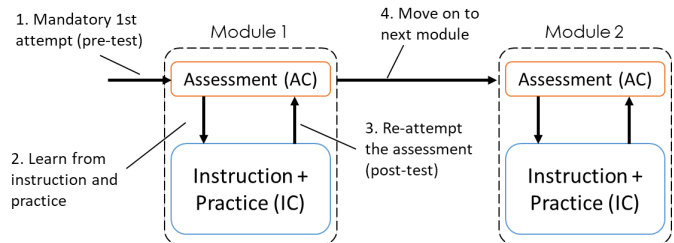

Figure 1: Schematic design of OLM and OLM sequence.

The IC of each module can contain a variety of learning resources, including text, figures, videos, solved examples, practice problems and simulations (Figure 2B). In the current study the content in the IC were limited to instructional text, figures and practice problems with immediate feedback. Accessing and interacting with the IC, including answering practice problems, was optional for passing the module and did not count toward any course credit.

Regardless of whether a student interacted with the IC, they could make a limited number of additional attempts on the AC following their initial attempt. Some students made more than one attempt on the AC before accessing the IC, and some did not access the IC at all. On each additional attempt, students received a new, isomorphic problem set that differed from the first set only by surface features. In the current implementation, students were given full credit if they passed the AC on any of the attempts.

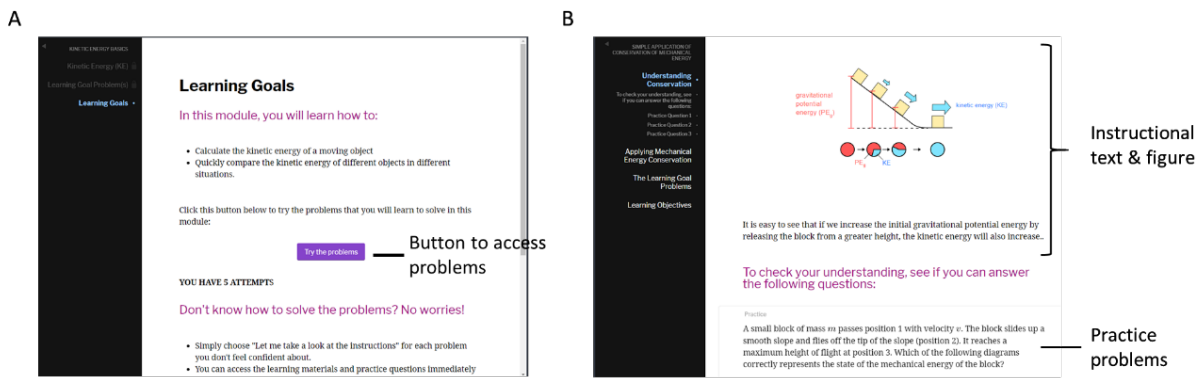

Figure 2: Screen capture of the OLM user interface. A: The first page upon accessing the module. Students must click a button to attempt the AC before having access to the IC. The links to IC on the left navigation bar (black) are initially locked (greyed out). B: A page in the IC containing both instructional figure and text and an optional practice problem.

By design, a student should only be allowed to proceed to the next OLM either after passing the current one, or after exhausting all available attempts on the AC. Due to limitations of the platform at the time of the study, the current implementation allowed a student to access the next module after making the first attempt on the AC of the current module, regardless of the outcome. However, students were not informed of this setting, and were strongly encouraged to finish each module (either passing or exhausting all attempts) before moving on to the next one. As shown in section 3.1, most of the students completed the modules in the given order, and the small number of "out of order" events were excluded from analysis.

## 2.2 Details of the OLM sequence in the current study

In the current study, the AC of each OLM contained 3 or 4 sets of isomorphic assessment problems, each set consisting of either 2 relatively simpler conceptual questions or 1 more complex calculation problem (see Table 1 and Appendix I), all of which were in multiple choice format. For each attempt at the AC, a student was presented with a problem set in a fixed order. If the student failed to solve all of the isomorphic problem sets, the system would return to the first set until no attempts were remaining. Each student was allowed 5 attempts on the AC of each module, providing them with the opportunity to attempt the $1^{st}$ (and on some modules the $2^{nd}$) problem set twice. The IC was locked from access during an AC attempt.

After submitting their answers to the problem set, students were informed which problems they answered correctly, but not the correct answer of the problems. A student would pass and receive credit for the module if they correctly answered all the questions in one problem set on any given attempt.

The IC consisted of both instructional text and practice problems divided into multiple pages, created by the author with the assistance of four undergraduate research assistants. Students received immediate feedback after attempting any practice problem and were provided with a complete solution on some of the harder practice problems. Each module's IC typically contained 2-3 pages and 2-5 practice problems.

The OLM sequence covered the topic of mechanical energy for calculus based introductory level college physics course. The title of each module is listed in Table 1:

Table 1: The title and assessment problem type of the ten OLMs assigned as homework in the current study.

| Modules | Title | Assessment Problem Type |
|---|---|---|
| 1 | Kinetic Energy | Conceptual |
| 2 | Work by a constant force | Conceptual |
| 3 | Work and Kinetic Energy | Numeric |
| 4 | Potential Energy | Conceptual |
| 5 | When is Mechanical Energy Conserved | Conceptual |
| 6 | Simple Application of Conservation of Mechanical Energy | Conceptual |
| 7 | Problems Using Conservation of Mechanical Energy | Both |
| 8 | Problems with Two Types of Potential Energy | Numeric |
| 9 | Mechanical Energy and Non-Conservative Work | Numeric |
| 10 | More mechanical energy problems | Numeric |

The first six modules focused on conceptual understanding of key concepts, and the AC for those modules consists of mostly conceptual questions (with the exception of module 3, Work and Kinetic Energy). Modules 7-10 focused on solving increasingly complex problems involving mechanical energy, and the ACs consist of numerical calculation problems (sample problems are presented in Appendix I). Many of the assessment problems were either modified from or inspired by research based conceptual surveys [40] and other earlier PER studies [5].

## 2.3 Implementation of the OLM sequence

The OLMs were implemented and hosted on the award-winning, open-source online learning objects platform, Obojobo, developed by the Learning System and Technology (LS&T) team at the Center for Distributed Learning at University of Central Florida [41–44]. The modules were combined into a sequence and assigned to students via the Canvas learning management system [45].

The OLM sequence was assigned to students as homework in a traditional lecture based first semester college introductory physics course at the University of Central Florida in Fall 2017 semester. Modules 1-6 were released 7 days before modules 7-10, and all 10 modules were due 16 days after the release of the first six modules. Completing all 10 modules was worth 9% of the total course score (two weeks' worth of normal homework credit), and each module was weighted equally. The modules were released concurrently with classroom lectures on the same topic. No other assignments were assigned to the students during this period.

## 3 Data Collection, Process and Analysis

## 3.1 Initial Processing of time-stamped log data.

Time-stamped log data on the following types of events were collected through the Obojobo platform for each student: Entering and exiting both the IC and the AC; Entering and exiting each page in the IC; Starting and finishing an attempt on the AC; Viewing a practice or assessment problem; Submitting an answer to either an assessment problem or a practice problem; Outcome of each attempt on the AC (number of problems answered correctly).

A small fraction of events that belong to the following two categories were excluded from the analysis in the current study:
1. Events that happened after a student passed a given module. (< 8% of total events)
2. Events on one module that happened after the student had started to interact with the next module. (< 3% of total events)

The reason for excluding both types of events is that it is difficult to determine how they affected students' performance on the AC, and most of those events took place after the assignment due date, presumably when students were reviewing those modules to prepare for an upcoming exam. In addition, due to a mistake in the settings of one module early on, 11 students accessed the IC of module 3 without making an initial attempt on the AC. The events from those 11 students on module 3 were also excluded from most of the analysis.

## 3.2 Attempts before and after learning from IC (pre and post testing)

Students' incoming knowledge and learning gains are measured by their performance (pass or fail) on attempts before and after accessing the IC. The precise definition of attempt before and after learning from the IC is as follows:

**Attempt before learning (ABL):** The first or second attempts on the AC before the student accessed the instructional components. Each attempt is marked by the events of a student entering and leaving the AC. In a few cases (less than 3% of all cases), a student will interact with the IC after more than 3 attempts on the AC. Those events were excluded from most analysis, since those students will have too few attempts left after interacting with the IC, leading to an inaccurate measure of their level of mastery after learning. Those students are categorized as "No Learning" (NL) on the particular module, together with students who never interacted with the IC at all.

**Attempt after learning (AAL):** Any attempt on the AC that occurred after the majority of learning events with the IC in an OLM has finished. See next section for the precise definition for "majority of learning events." In the current study, since the AC of each module contains 3-4 sets of isomorphic problems but allows for 5 attempts, a student is considered as "fail" if they did not pass the module after attempting each isomorphic problem once, even though some of those students did answer the assessment problems correctly on their $4^{th}$ or $5^{th}$ attempt.

The time spent on each assessment attempt is recorded as the time between the start of the assessment, and when the student clicks the submit button on the assessment, regardless of the number of problems in the AC.

### 3.3 Interaction with the IC

**Major Learning Session (MLS):** We will refer to a single instance of continuous interaction with the IC from a student as a learning sessions (LS). A learning session starts when the student navigates into the IC and ends when the student navigates away from the IC, either entering the AC to start an attempt or exiting the module (for example, by closing the browser). In 93% of the cases, each student had only one LS on the IC of any given module, which is labeled as the MLS of the student. In 6% of the cases, a student had a second LS that was longer than 60 seconds and at least 30% as long as their longest LS. In those cases, the two LSs were combined into one single MLS, and the attempt in between neglected from analysis. In less than 1% of the cases did we observe a student making more than two LSs longer than 60 seconds. In those cases, only the two longest LSs were combined into the MLS for that student on that module. Multiple different criterion for identifying a second longest LS were tested, with little impact on the outcome. In this study we assumed that the MLS captures most of a student's learning from the IC. Therefore, AAL in section 3.2 is defined as attempts made after a student's MLS.

The MLS divides a student's interaction with each OLM into three stages: The *pre-learning stage*, in which the student makes all ABLs before staring the MLS, the *learning stage*, which is the MLS, and in some occasions an AC attempt in between, and the *post-learning stage*, which contains all the AALs and can sometime contain a brief second interaction with the IC between attempts. A schematic illustration of the different stages is presented in Figure 3.

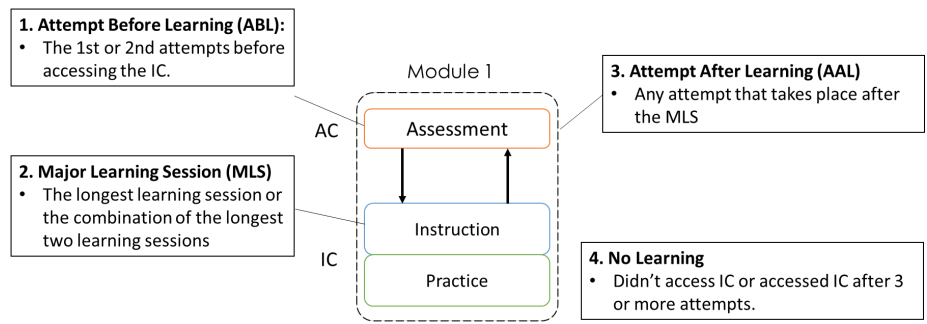

Figure 3: Schematic illustration of the actions and outcomes considered in the current study for a single OLM module.

### 3.4 Data analysis:

All the data analysis in the current study was conducted in R [46] using the "tidyverse" package [47]. Multi-component normal distribution fits of duration data were conducted using the package "mixtools"[48]. Visualizations (sunburst charts) of data analysis results were created using D3[49].

# 4 Results

## 4.1 Number of students that attempted and passed each OLM:

A total of 230 students attempted at least 1 module in the sequence. The number of students who made at least one attempt on each module, and who passed each module on either ABL or AAL, are listed in Table 2. The last row lists the size of the NL group on each module, which is defined in section 3.2.

Table 2: Number of students who attempted each module, passed each module on ABL or AAL, and number of students who are categorized as "No Learning" (NL).

|  | module 1 | module 2 | module 3 | module 4 | module 5 | module 6 | module 7 | module 8 | module 9 | module 10 |
|---|---|---|---|---|---|---|---|---|---|---|
| N | 230 | 230 | 230 | 230 | 229 | 229 | 227 | 225 | 223 | 223 |
| Pass (ABL) | 63 | 39 | 118 | 47 | 21 | 88 | 28 | 102 | 89 | 54 |
| Pass (AAL) | 105 | 120 | 74 | 85 | 86 | 112 | 30 | 55 | 85 | 53 |
| NL | 19 | 27 | 9 | 31 | 28 | 23 | 28 | 31 | 37 | 58 |

We emphasize again that since the OLMs were assigned as homework, there can be substantial variations in students' test-taking effort during each attempt on each module, as well as in the level of engagement with the instructional materials. Therefore, the number of passing students listed in Table 2 cannot directly provide accurate information on the effectiveness of the module, before information on students' test-taking effort and learning effort are being taken into account, which is described in detail in the next two sections.

## 4.2 Students' test-taking effort on ABL and AAL

Our analysis of student test-taking effort focused on identifying two types of test-taking behavior on the AC of each module: abnormally low test-taking effort on both ABL and AAL, and higher than normal test-taking effort on AAL for more challenging problems.

**Low Test-taking Effort (Brief):** Abnormally low test-taking effort is characterized by a very short duration between start and end of an assessment. Many earlier studies have shown that such short duration is frequently associated with either guessing [15,16] or answer copying [23,28,50]. In particular, when low test-taking effort is accompanied by high accuracy, it could indicate that either students are copying the correct answer from another source, or the correct answer in a multiple-choice problem is too easy to guess.

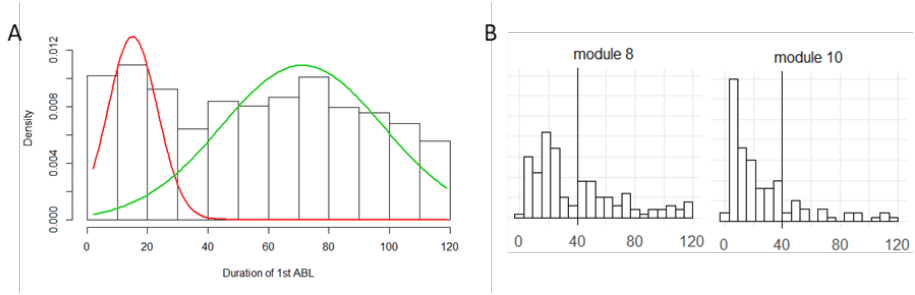

Figure 4: Distribution of ABL duration under 120 seconds and best fit mixture model. A: Distribution of all ABL duration and 2-component best fit model. B: Distribution of ABL duration for modules 8 and 10, which contain a substantial number of short duration attempts.

We determine the cut-off time for low test-taking effort on ABL via a mixture-model analysis technique described in detail in two previous publications [25,43]. The method fits the distribution of time-on-task data using multi-component normal distributions. For the current analysis, we choose to combine data from students' first attempts on all 10 modules, and fit the distribution of problem-solving duration under 120 seconds using a 2-component normal distribution, as shown in Figure 4A.

There are three reasons for combining data from all 10 modules to obtain a uniform cutoff. First, the assessment of each module contains either 1 or 2 problems. Making a random guess on those problems should take similar amount of time on each module. Second, on several modules there were too few students with short problem-solving duration to accurately identify a low test-taking effort group using mixture models. Finally, since low test-taking efforts are more likely associated with random guessing or answer copying, the duration of those attempts should be largely independent of the actual problems presented. In fact, we found significantly more students with short problem-solving duration on complicated calculation problems, which should take longer to solve, compared to easier conceptual problems (see Figure 9).

The two components of the best fit model intersected at around 30 seconds, and the first component (red) decreases to below 5% of its peak value at 40s. Some previous studies on online physics homework have suggested longer cutoffs (60s) for identifying "guessing" or "answer copying" behavior [23,28,43]. Furthermore, on modules with more short-duration attempts, such as modules 8 and 10, there seems to be a natural gap at 40s in the time-on-task profile (Figure 4B). Therefore, we use 40s as a uniform cutoff for identifying the low test-taking effort population. We will refer to attempts shorter than 40s as "Brief" attempts, and the rest as "Normal" attempts. Empirically, 40s is roughly the time that it takes to load the assessment page and quickly read through the problem body. (Distributions of all ABL duration under 120s can be found in Appendix II).

For assessment attempts after learning (AAL), data from 2 modules must be excluded before the same analysis can be performed. In Figure 5, we plot the histogram of the duration on the AAL of each module on log scale. Students who passed the assessment are plotted in green and those who failed the assessment are plotted in red. For those who failed, duration of their first post-learning attempt was used. For those who passed, the duration of their passing attempt was used. From the figure it is clear that, for modules 2 and 6, almost half of the students completed the attempt in under 40 seconds (the leftmost vertical black line). In contrast, on these same modules less than 10% of the students completed their first attempt before learning under 40 seconds. A possible explanation is that both modules involve conceptual problems that do not require calculation, and the isomorphic problems presented on each attempt are very similar. Therefore, students do not need to fully read the entire problem body again after the first attempt and can submit an answer very quickly.

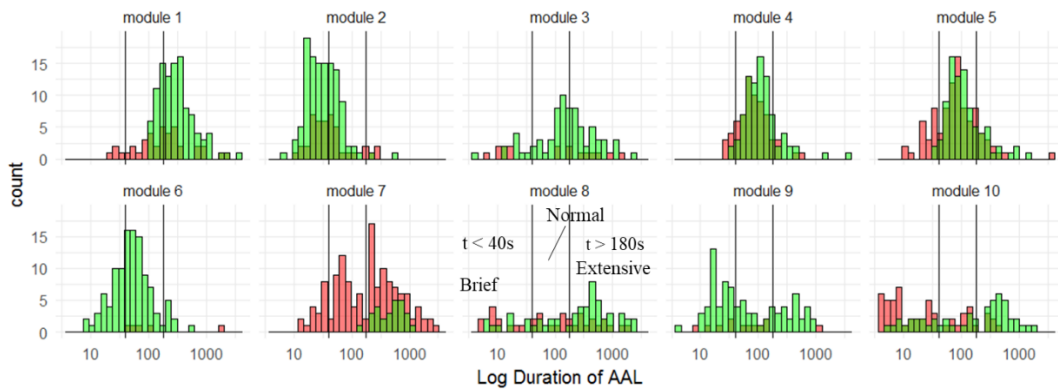

Figure 5: Distribution of log transformed durations of students' attempt after learning (AAL). Passed attempts where colored in green and failed attempts were colored in red

Conducting the same mixture model analysis on AAL duration data after excluding modules 2 and 6 yields almost identical result, with the two normal components intersecting between 30 and 40 seconds (data not shown). Therefore, for AAL on all modules except 2 and 6, an attempt < 40s is labeled as "Brief", and all other attempts are labeled "Normal". On modules 2 and 6, all AAL attempts were labeled as "Normal".

**High Test-taking Effort (Extensive):** From Figure 5, it can be seen that for modules 1-6, the duration distribution is similar to a single normal distribution, whereas on modules 7-10, a second peak in the distribution can be clearly observed for duration longer than 180 seconds (vertical black line on the right). In addition, attempts longer than180s have significantly higher passing rates ($p < 0.01$, Fisher's Exact Test) than those less than 180s (except for module 9, see Appendix II for complete statistical test results).

Because the ACs for modules 7-10 involve more complicated numerical calculation problems, it is possible that students who spent >180 seconds on solving those problems are more likely than others to be using the correct method. Therefore, on those four modules, we will label attempts longer than 180s as "Extensive". For attempts before learning, there seems to be a similar trend but much less identifiable, so for the current study we will not identify "Extensive" attempts for ABLs on the last four modules.

**Combining assessment outcomes with test-taking effort.** Combining the low, normal, and high test-taking efforts with the two assessment outcomes (pass/fail) results in six types of assessment performance for the ABL and AAL of each module: Brief-Fail (BF), Brief-Pass (BP), Normal-Fail (NF), Normal-Pass (NP), Extensive-Fail (EF), Extensive-Pass (EP). The six categories are illustrated in Figure 8B, in which the color corresponds to outcome and the darkness represents test-taking effort. Note that the last two types, EP and EF, are only applicable to the post-learning attempts on modules 7-10. The fraction of students belonging to each type on each module are plotted on the inner and outer rings of the sunburst charts in Figure 9 and explained in detail in section 4.4.

### 4.3 Analysis of Student Learning Effort

Information about student learning effort on each module was obtained through analysis of the duration of each student's major learning session (MLS). Although the actual relation between time-on-task and learning effort is likely much more complex, our basic hypothesis is that qualitatively different learning behaviors, such as quickly skimming through the instruction vs. carefully reading the instructional material will likely take very different amounts of time, and can be identified by clustering analysis techniques such as mixture-models [25]. While several other variables, such as number of practice problems attempted or average attempt on each practice problem may also be related to learning effort, our previous study on OLMs showed that time-on-task is the most sensitive variable for identifying different levels of learning effort [25].

Unlike the analysis for test-taking effort, uniform cut-offs for all modules are undesirable for learning effort for two reasons. First, the amount of content in the IC of each module can be quite different. Second, since accessing the IC is not required for passing the module, students have little incentive to open the IC and quickly close it. Those very brief events are also not meaningful for analysis. Therefore, the MLS duration of each module needs to be analyzed individually.

We plot the histogram of MLS duration for the 10 modules on log scale in Figure 6A. To identify different types of learning behavior for each module, we fit the log duration distribution of each module with a multi-component normal distribution. The optimum number of normal components is determined by a parametric bootstrapping algorithm comparing likelihood of the *k*- component fit to the *k*+1 component fit over 2000 iterations [48]. The results are shown in Table 3. The *p*-values indicate whether the log likelihood of the *k*+1 component fit is significantly more favorable than that of the *n*-component fit.

As shown in the table, the duration distribution of modules 1, 2, 4, 5, 6, and 9 are best fitted with a single normal distribution, modules 3, 7, and 8 are best fitted by 2-component models, and module 10 is best fitted using a 3-component model. This result is also in agreement with a Shapiro-Wilk normality test on the distributions, which reveals that modules 3 and 7-10 are significantly non-normal. (see Appendix II for statistical test results).

Table 3: Outcome of parametric bootstrap algorithm comparing the likelihood of *k* component fit vs. *k*+1 component fit. *p*-values indicate the likelihood of *k*+1 component fit being significantly higher than that of a *k* component fit. The "Comp." row contains the optimum number of components recommended by the algorithm.

|       | module 1 | module 2 | module 3 | module 4 | module 5 | module 6 | module 7 | module 8 | module 9 | module 10 |
|-------|----------|----------|----------|----------|----------|----------|----------|----------|----------|-----------|
| p1-2  | 0.29     | 0.48     | 0.04*    | 0.24     | 0.9      | 0.68     | 0.02*    | 0.01*    | 0.28     | 0.00**    |
| p2-3  |          |          | 0.35     |          |          |          | 0.45     | 0.18     |          | 0.03*     |
| p3-4  |          |          |          |          |          |          |          |          |          | 0.89      |
| Comp. | 1        | 1        | 2        | 1        | 1        | 1        | 2        | 2        | 1        | 3         |

For modules 3 and 8, the two components of the best fit model consist of a smaller normal component with less mean duration and a larger normal component with longer mean duration, as is shown in Figure 6A. For those two modules, we will refer to the smaller component as "Brief" learning, and the larger component as "Normal" learning. The cut-off point is set as the point of intersection between the two components. Note that "Brief" learning does not necessarily imply insufficient or inadequate engagement with the instructional material for each module.

In the 2-component best fit model for module 7, the smaller normal component captured a small group of students with longer than normal learning duration, whereas the larger normal component captured all the other students. However, the 3-component best fit model, which fits the duration profile equally well as the 2-component model (see Table 3), identified a "Brief" learning component, as shown in Figure 6B. For this study, we will combine the two clusters with longer mean durations into one "Normal" learning cluster, to keep the categorization uniform across all modules and simplify the outcome.

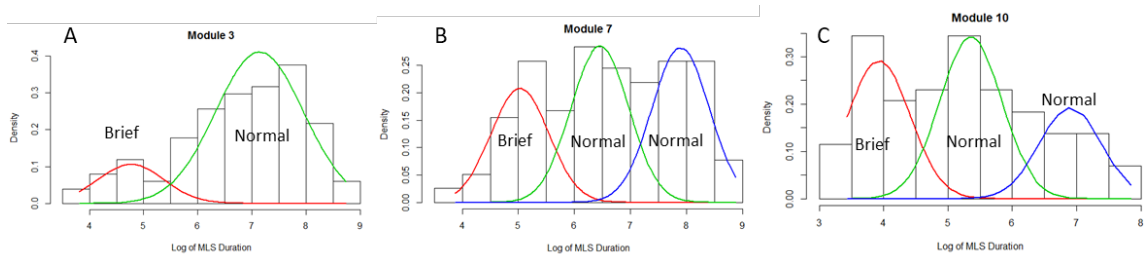

Figure 6: Mixture-model best fit for modules 3 (2-component), 7 and 10 (3-components). For module 7, a 2-component model fits the distribution equally well as a 3-component fit.

For module 10, the best fit model contains three components with increasing mean duration. For the same reason as above, we will combine the two longer components into one "Normal" learning cluster, as shown in Figure 6C.

For modules that are best fit with a 1-component normal model, we will categorize students as "Brief" if the log of their MLS duration is 1 standard deviation below the population mean, following a similar approach in an earlier study [25]. The cut-off between "Brief" and "Normal" for all modules are shown as vertical black lines in Figure 7, and the values listed in Appendix II.

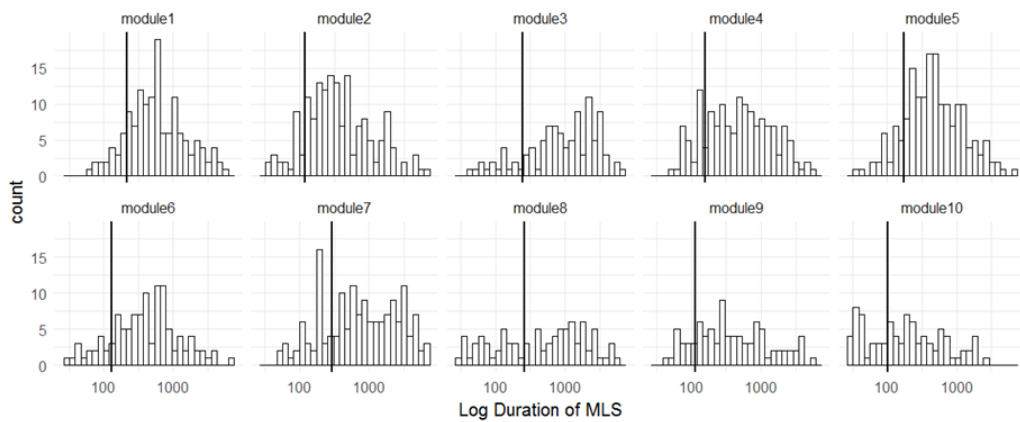

Figure 7: Distributions of log-transformed duration of MLS for each module. The vertical line is the cutoff between Brief and Normal learning events determined by either mixture-model analysis or one standard deviation below the population mean.

In section 4.5, we will analyze the extent to which the "Brief" and "Normal" learning clusters identified for each module correlates with different learning outcomes and test-taking effort on post-learning attempts (AAL). In section 5.3, we will discuss the validity of the selection of cut-off values between different clusters.

## 4.4 Visualizing Assessment Outcome, Learning Effort and Test-taking Effort

To synthesize and visualize students' assessment outcomes, test-taking effort, and learning effort on each module, we plot information for each module on a sunburst chart, explained in Figure 8 and displayed in full in Figure 9. First, the fraction of students having each of the six types of attempt behavior (see section 4.2) on ABL is plotted on the inner ring (Figure 8B). For students who failed on ABL, the distribution of "Brief" (light yellow) and "Normal" (darker yellow) learning efforts following ABL are plotted on the middle ring (Figure 8C). Finally, the distribution of student assessment performance on attempts after learning (AAL) are plotted on the outer ring, grouped by their learning efforts. In this representation, each student moves from the inner ring to the outer ring during the interaction with an OLM. Students who passed the assessment on ABL do not have either learning event or AAL attempts on that module, resulting in two gaps in the middle and outer ring for each chart.

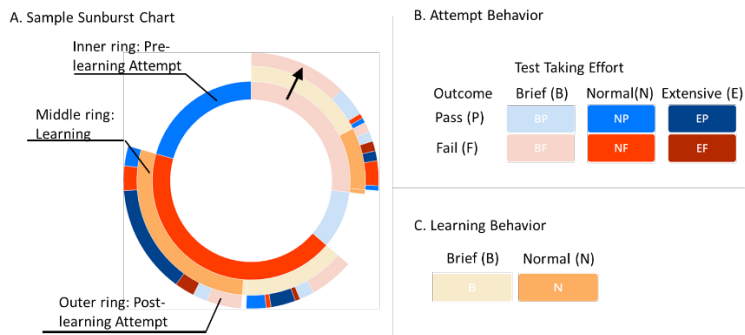

Figure 8: A: Example sunburst chart for one OLM. B: Color scheme for different attempt behavior groups. C. Color scheme for different learning behavior groups.

Figure 9 shows the sunburst charts for all 10 modules. A notable feature is that even though all of the modules are assigned to students over a period of 2 weeks, and many students completed multiple modules in one day, there are still considerable variations in student performance and behavior on each module. In section 5, we discuss in detail how these charts can provide instructors with useful information on the effectiveness of instructional materials in each module, and suggest directions for future improvements.

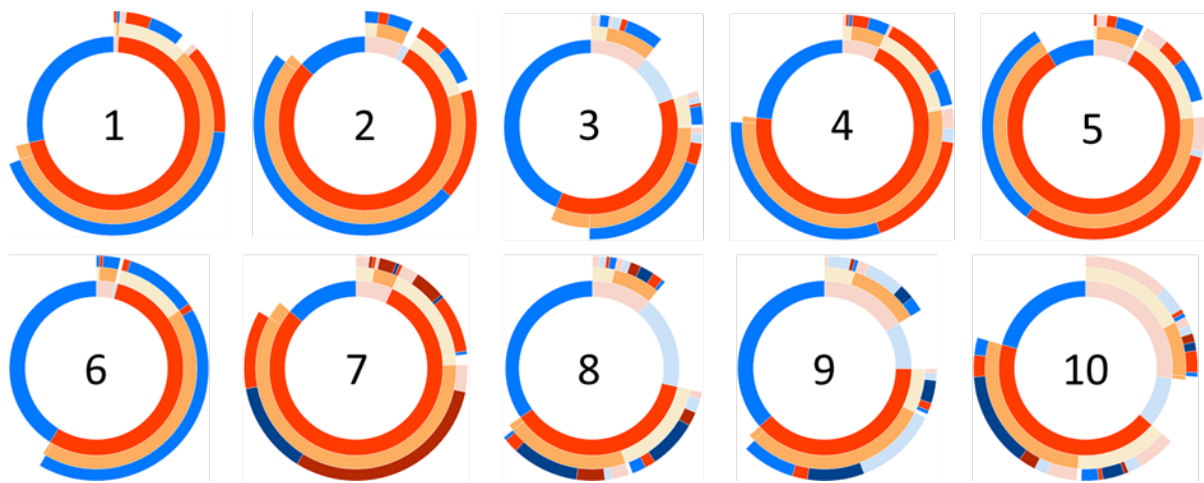

Figure 9: Sunburst chart representation of students' learning behavior and engagement for each of the 10 OLMs. Some of the small gaps on the outer ring (post-learning attempts) are caused by a few students not completing each module in the given order, therefore their data is excluded from further analysis.

### 4.5 Correlations between Learning Effort and Assessment Behaviors

Having divided students' learning efforts into "Brief" and "Normal" for each module in section 4.3, a natural follow up question is whether the two types of learning efforts are correlated with different learning outcomes on each module. More specifically, we ask the question:

*Q1:* Are students' outcomes (pass/fail) on post-learning attempts (AAL) correlated with their learning efforts (Brief or Normal) prior to the attempts?

To obtain a more accurate answer to this question, we need to take students' test-taking effort into consideration by answering the following two additional questions:

*Q2:* Are students' test-taking efforts on *pre-learning attempts (ABL)* correlated with their learning effort?

*Q3:* Are students' test-taking efforts on *post-learning attempts (AAL)* correlated with their learning effort?

Q2 is important because different types if test-taking effortson pre-learning attempts could be an indication of different learning behavior that took a similar amount of time. For example, a student who was a "Normal" problem solver on the pre-learning attempt may be more likely to become a "Brief" learner because he/she had some initial understanding of the content and only needed a quick refresh or a reminder, rather than being unmotivated or unengaged. Therefore, if ABL test-taking efforts and learning efforts are uncorrelated on a given module, it is worthwhile to divide the student population according to a combination of both measures.

Q3 is important because for modules with assessment problems that are either too challenging or too easy, the passing percentage may not reflect the difference in a student's level of skill mastery after learning. In that case, the student's test-taking effort on post-learning attempts may provide some information on the student's level of mastery. For example, it is possible that a student who spent an adequate amount of time solving the problem is more likely to have better understanding than a student who made random guesses, even though both answered the question incorrectly.

In Table 4, we summarize the main results (*p*-values of multiple Fisher's exact tests) of correlations between students' learning efforts with assessment outcomes and test-taking efforts to answer Q1-Q3. Additional test statistics are listed in Appendix II.

Table 4: *p*-values for Fisher's exact test on correlations between learning effort (MLS) and ABL test-taking effort (row 1), AAL outcome (row 2) and AAL test-taking effort (rows 3 and 4). In rows 2-4, the student population is divided into four Engagement Modes (EM). In row 4, AAL Extensive group are only identified for the last four modules.

| Types of Correlation | 1 | 2 | 3 | 4 | 5 | 6 | 7 | 8 | 9 | 10 |
|---|---|---|---|---|---|---|---|---|---|---|
| MLS-ABL(Brief) | 0.31 | 0.24 | 0.23 | 1.00 | 0.48 | 1.00 | 0.10 | 0.46 | 0.29 | 0.00* |
| EM – AAL (pass) | 0.00* | 0.06 | 0.88 | 0.01* | 0.62 | 0.44 | 0.1 | 0.34 | 0.13 | 0.01* |
| EM – AAL (Brief) | 0.15 | 0.19 | 0.01* | 0.22 | 0.04* | 0.3 | 0.00** | 0.16 | 0.04* | 0.00** |
| EM – AAL (Extensive) | NA | NA | NA | NA | NA | NA | 0.00** | 0.00* | 0.21 | 0.00* |

Regarding Q2, the first row in Table 4 shows that on 9 out of 10 modules, students' test-taking effort on pre-learning attempts are not correlated with their learning efforts (MLS). Therefore, when answering Q1 and Q3, we will divide the student population into four "Engagement Modes" (EM) by combining ABL test-taking effort with learning effort: *Normal-Normal (NN), Normal-Brief (NB), Brief-Normal (BN), Brief-Brief (BB)*. The first letter signifies ABL test-taking effort and the second letter represents the learning effort. The number of students in each mode on each module can also be found in Appendix II.

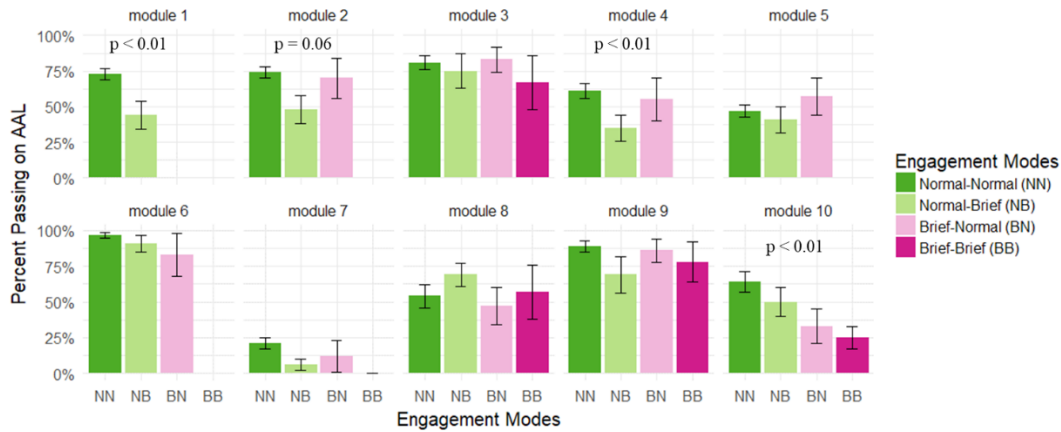

Figure 10: Percentage of students passing on AAL for each of the four engagement modes. Modes with < 5 students are neglected from analysis.

To answer Q1, the *p*-values (Fisher's exact test) for correlation between passing on AAL and the four EMs is listed in row 2 of Table 4, and the fraction of passing students in each EM plotted in Figure 10. EM modes with less than 5 students are excluded from analysis. On modules 1, 2, and 4, the "NB" modes are significantly correlated with lower AAL passing percentages, whereas the "BB" modes are removed because of low student number. On module 10, both "BN" and "BB" modes have significantly lower passing percentage than the "NN" mode. In other words,

Brief problem solvers on pre-learning attempts are more likely to fail on post-learning attempts in module 10, regardless of their learning effort.

For Q3, we first examine the correlation between the 4 EMs and "Brief" problem solving on AAL. On five modules (3, 5, 7, 9, 10), the correlation is statistically significant (row 3 of Table 4). As shown in Figure 11, on each module except module 5, the BB mode has significantly higher fraction of "Brief" problem solvers on AAL than all other modes. On module 9, the BN mode also has significantly more "Brief" problem solvers on AAL. Notice that on all those modules, the passing percentage is similar among all four EMs. On the other hand, on modules 1, 2, and 4, which had significant difference in post-learning passing rates, there were few post-learning "Brief" problem solvers. It is possible that on modules 3, 7, 9, and 10, more students passed the assessment by guessing or answer copying, which resulted in the "BB" mode having similar passing rate as the other three modes.

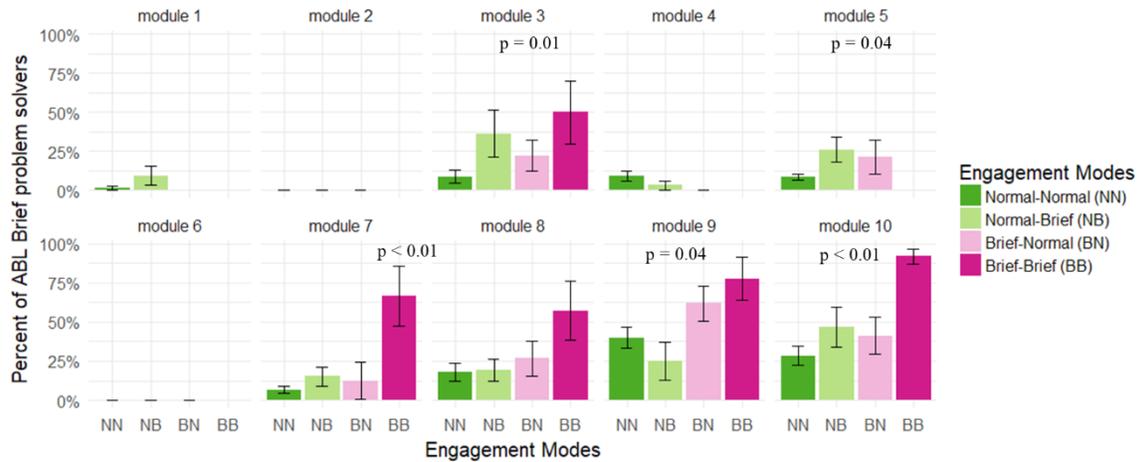

Figure 11: Fraction of AAL Brief problem solvers among the four engagement modes. Modes with < 5 students are neglected. Note that all students were categorized as Normal on AAL attempts for modules 2 and 6.

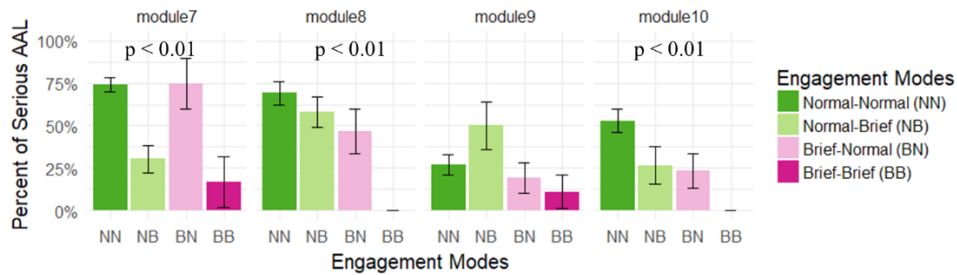

Figure 12: Fraction of "Extensive" problem solvers in modules 7 – 10 among the four Engagement Modes.

In addition, we also examined the correlation between the 4 EMs and "Extensive" AAL problem solvers on modules 7-10. As shown in row 4 of Table 4 and Figure 12, on all modules the "BB" mode had significantly less Extensive problem solvers on AAL. An interesting observation is that while on module 7, the "NB" mode is still significantly less likely to contain "Extensive" problem solvers, on modules 8 and 9 the "NB" modes are equally as likely, if not more likely to contain Extensive problem solvers. One possible reason is that the "Brief" learning population in the "NB" mode on those two modules mainly consisted of students who had some level of understanding of the content covered in the modules, and only needed a quick reminder or refresh to pass the assessment. The large difference between "BB" and "NB" modes in terms of AAL test-taking efforts on those modules also supports the hypothesis that different test-taking effort on ABL may indicate different causes for similar "Brief" learning behavior.

In all of the analysis in this section, a "hybrid" Fisher's exact test was conducted when the contingency table contains occasional small expected counts. For 2 by 4 contingency tables (rows 2-4 of Table 4), Fisher's test was conducted with Freeman-Halton extension. Error bars on Figures 10-12 represent binomial error.

## 5 Discussion:

### 5.1 Information on the Effectiveness of Each OLM

Here we will list the most prominent features that can be easily identified from the sunburst charts in Figure 9, and discuss their implications for the effectiveness of the OLMs. Statistical test results from section 4.5 will be used to support some of the implications.

1. On module 7, most "Normal" learners failed to pass the module after learning. Even "Extensive" problem solvers on post-learning attempts mostly failed to pass, as can be seen from the dark red band on the outer ring. This suggests that the learning resources in module 7 are inadequate for helping most students correctly answer the assessment problems, even for those who were adequately engaged with the resources. The chart for module 5 shows a similar pattern to a lesser extent.

2. On modules 3, 8, 9, and 10, a larger fraction of "Brief" problem solvers passed the module on their pre-learning attempts, which is represented by a longer light blue band on the inner ring and a significant gap in the middle and outer ring opening to the right. There can be at least two possible explanations. First, there may not be enough good distractors for the multiple-choice assessment problems, which enabled students to easily guess the correct answers. Second, some students may have found those problems challenging and copied the answer from a peer. Both explanations could also account for the high post-learning passing rate and low test-taking efforts observed for students in the "BB" engagement mode described in section 4.5.

   On those modules, most students in the "NN" mode (red inner ring with dark-yellow middle ring) passed the assessment as either "Normal" (blue) or "Extensive" (dark blue) problem solver on the post-learning attempts (outer ring), which suggests that the instructional materials are adequate at least for those students who fully engaged with the learning resources.

3. For module 10, a much larger fraction of students briefly interacted with both the AC and the IC and failed to pass the module, as shown by the large pink, light-yellow, pink band on the top right of the chart ("Brief-Brief-Brief"), unique to module 10. This pattern suggests that a larger than normal number of students are not fully engaged with the module, either because it is the last module, or because it is the most challenging module in the sequence. Unlike module 7, the low overall passing rate on module 10 is more related to a lack of student engagement.

4. Almost all students passed module 6, regardless of their learning effort, seen from the predominantly blue outer ring, and the long blue band on the inner ring. Notice that there were only a small number of "Brief" problem solvers on the pre-learning attempt, and almost none of them passed on those attempts. Therefore, it is likely that the assessment problems in module 6 are easy to solve after learning, rather than being easy to guess.

5. On modules 1, 2 and 4, a good fraction of "Normal" learners passed the module after learning. More importantly, "Brief" learners were significantly less likely to pass after learning, as confirmed by correlation analysis in section 4.5. In addition, on the pre-learning attempts there were essentially no "Brief" problem solvers who passed (no light blue band on the inner ring). Those results suggest that the instructional resources in these modules are effective, and the assessment problems have adequate discrimination power.

6. In contrast, on modules 8 and 9, students in the NB mode, i.e. Brief learners who were Normal problem solvers on pre-learning attempts, are equally if not more likely to pass on post-learning attempts as "Normal" or "Extensive" problem solvers, as confirmed by statistical test results in section 4.5. One possible explanation is that students learned how to solve the problems from a source other than the IC of the OLMs. However, given the fact that that modules 7-10 are all about applications of the conservation of mechanical energy principle to problems at increasing levels of complexity, it is also plausible that some students have learned the concept from module 7 and only required a quick reminder to successfully solve problems on modules 8 and 9, which could also have contributed to the high pre-learning passing rate on those two modules.

## 5.2 Unique Advantages of the OLM design

As discussed above, the mastery learning design of the OLMs allows information about the effectiveness of each module to be extracted from data on students' interactions with the modules themselves. More importantly, the analysis results also provide some clear suggestions for improving the OLM sequence:

1. For modules 7 and 5, improve the quality of instructional materials.
2. For modules 3, 8, and 9, improvements should first focus on the quality of assessment problems, such as adding more distractors.
3. Module 6 needs to have more challenging assessment problems.
4. For module 10, the results suggest a significant increase in disengagement among students. Therefore, students' learning outcomes from module 10 might benefit more from strategies that improve engagement.

Finally, the results also suggest that modules 1, 2, and 4 do not need to be prioritized for future improvements. These suggestions can be highly valuable for instructors who only have a limited amount of time and resources to make incremental improvements to instruction each semester, directing their efforts towards the most needed improvements. More importantly, applying the same analysis methods on improved modules could reveal the effectiveness of new improvements, which can help advance both instructional practice and education research.

Few other instructional designs or assessment methods can provide this type of high-resolution information on the quality of instructional resources. For example, while many online learning platforms such as Canvas [45] provide usage data such as page views and performance data such as quiz scores, it is difficult to make meaningful connections between the two, since in most conventional online courses learning materials and assessments are separated further apart. As a result, it is much harder to determine whether students viewing (for example) 20 pages and 5 videos over the period of a week is related to their average score on a weekly quiz.

On the other hand, conventional pre/post tests and clinical experiments cannot be conducted at such a high frequency in a real classroom. Pre/post testing also cannot provide information on students' test-taking and learning efforts, both of which are critical in the current analysis. While clinical experiments can measure behavior such as guessing or brief learning with higher accuracy, it would be impossible to know the fraction of students who would display this type of behavior in an authentic learning setting, such as the disengaged student population on module 10.

Therefore, with their unique advantages, OLMs can serve as an important supplement to the existing methods for quickly and accurately measuring the effectiveness of learning resources.

## 5.3 Validity of Clustering method

Most of the results in the current analysis were obtained through a clustering algorithm (mixture model) that divides a continuous time distribution into multiple clusters. To what extent do the cutoffs between two clusters correspond to actual differences in student behavior, and how do some of the simplifications, such as combining the "Extensive" and "Normal" learning clusters in two modules, affect the results need to be examined by multiple carefully designed follow up studies in the future. Here, we present some evidence that supports the validity of some of the cutoffs used in the study.

For the 40s cutoff that divided the "Brief" and "Normal" problem solving clusters, it is noteworthy that on the two modules (2 and 6) where a large fraction of students answered in under 40s on attempts after learning, there were very few students who answered in < 40s on pre-learning attempts. On the contrary, pre-learning attempts < 40s significantly increased on modules 8-10, which contain complex numerical calculation questions. Therefore, it seems that most < 40s answers on the pre-learning attempts and most of the post-learning attempts were instances of guessing or answer copying, rather than very quick problem solving.

For the cutoffs that divided the "Brief" and "Normal" learning clusters in each module, we tested two reasonable alternative methods for determining those cutoffs. Under the two alternative sets of cutoffs, most of the correlations between the four engagement modes and learning outcomes remain the same, with only one or two exceptions. The detailed results and discussion can be found in Appendix II.

## 5.4 Caveats and Future Directions.

Finally, we will address several caveats in the current study and suggest some possible directions for future studies.

First, in this paper we used a very narrow definition for the "effectiveness" of learning resources: the increase in number of students passing each module before and after learning. A more comprehensive definition could include other important factors such as retention, transfer, and change in students' motivation and attitude. All of those factors can be examined in future studies involving specifically designed OLM sequences. For example, in a separate study we examined students' ability to transfer their learned skill to new situations using a sequence of three closely related OLMs [51]. Retention of learning can be assessed by, for example, students' performance on related problems in a proctored classroom exam following the completion of OLMs.

Secondly, in this study the ICs in each OLM contain instructional resources created by the first author, which are well aligned with the assessments in each module. Whether OLMs are able to provide information on the effectiveness of open educational resources online, such as the vast number of YouTube videos, is a more valuable question that will be answered in future studies.

Thirdly, most of the analysis in the current study focused on distinguishing between "Brief" and "Normal" learning or test-taking behavior. "Extensive" problem solving on all pre-learning attempts and "Extensive" learning identified for modules 7 and 10 are not considered in the current analysis. Analysis of "Extensive" behavior in future studies could yield more insight into both the quality of learning resources and student learning behavior in an online environment.

Finally, the current analysis focused on student behavior in each individual module. An important and potentially fruitful future research direction is to understand how individual student's behavior changes across all 10 modules. For example, how many "Extensive" learners on module 7 can pass modules 8 and 9 before learning or after only brief learning, by transferring their understanding to a new problem? When do some students start to disengage from the learning process? Tracing each student's behavior across all 10 modules could provide much insight into students' self-regulation and motivation in an online learning environment.

# 6 References

[1] Chen Z and Gladding G 2014 How to make a good animation: A grounded cognition model of how visual representation design affects the construction of abstract physics knowledge *Phys. Rev. Spec. Top. - Phys. Educ. Res.* **10** 010111

[2] Stelzer T, Gladding G, Mestre J and Brookes D T 2009 Comparing the efficacy of multimedia modules with traditional textbooks for learning introductory physics content *Am. J. Phys.* **77** 184

[3] Gladding G, Gutmann B, Schroeder N and Stelzer T 2015 Clinical study of student learning using mastery style versus immediate feedback online activities *Phys. Rev. Spec. Top. - Phys. Educ. Res.* **11** 1–8

[4] DeVore S, Marshman E and Singh C 2017 Challenge of engaging all students via self-paced interactive electronic learning tutorials for introductory physics *Phys. Rev. Phys. Educ. Res.* **13** 010127

[5] Fakcharoenphol W, Potter E and Stelzer T 2011 What students learn when studying physics practice exam problems *Phys. Rev. Spec. Top. - Phys. Educ. Res.* **7** 1–7

[6] Marshman E M, Devore S, Singh C, Slotta J D and Marshman E M 2018 Challenge of Helping Introductory Physics Students Transfer Their Learning by Engaging with a Self-Paced Learning Tutorial **5** 1–15

[7] Hake R R 1998 Interactive-engagement versus traditional methods: A six-thousand-student survey of mechanics test data for introductory physics courses *Am. J. Phys.* **66** 64–74

[8] Ding L, Reay N W, Lee A and Bao L 2008 Effects of testing conditions on conceptual survey results *Phys. Rev. Spec. Top. - Phys. Educ. Res.* **4** 2–7

[9] Adams W K and Wieman C E 2011 Development and Validation of Instruments to Measure Learning of Expert-Like Thinking *Int. J. Sci. Educ.* **33** 1289–312

[10] Dori Y J and Belcher J 2005 How Does Technology-Enabled Active Learning Affect Undergraduate Students' Understanding of Electromagnetism Concepts? *J. Learn. Sci.* **14** 243–79

[11] Hestenes D, Wells M and Swackhamer G 1992 Force concept inventory *Phys. Teach.* **30** 141

[12] Marx J D and Cummings K 2007 Normalized change *Am. J. Phys.* **75** 87–91

[13] Bao L 2006 Theoretical comparisons of average normalized gain calculations *Am. J. Phys.* **74** 917–22

[14] Schnipke D L and Scrams D J 1999 Modeling item response times with a two-state mixture model- a new approach to measuring speededness *J. Educ. Meas.* **34** 213–32

[15] Wise S L and Kong X 2005 Response Time Effort: A New Measure of Examinee Motivation in Computer-Based Tests *Appl. Meas. Educ.* **18** 163–83

[16] Barry C L, Horst S J, Finney S J, Brown A R and Kopp J P 2010 Do Examinees Have Similar Test-Taking Effort? A High-Stakes Question for Low-Stakes Testing *Int. J. Test.* **10** 342–63

[17] Davis D, Chen G, Hauff C and Houben G 2016 Gauging MOOC Learners' Adherence to

[18]  Liu Z, Brown R, Lynch C, Barnes T, Baker R, Bergner Y and McNamara D 2016 MOOC Learner Behaviors by Country and Culture; an Exploratory Analysis *Proceedings of the 9th International Conference on Educational Data Mining* pp 127–34
[19]  Kizilcec R F, Piech C and Schneider E 2013 Deconstructing Disengagement : Analyzing Learner Subpopulations in Massive Open Online Courses *Lak '13* 10
[20]  An T-S, Krauss C and Merceron A 2017 Can typical behaviors identified in MOOCs be discovered in other courses? *Proceedings of the 10th International Conference on Educational Data Mining* pp 220–5
[21]  Miyamoto Y R, Coleman C A, Williams J J, Whitehill J, Nesterko S and Reich J 2015 Beyond time-on-task: The relationship between spaced study and certification in MOOCs. *J. Learn. Anal.* **2** 47–69
[22]  Breslow L, Pritchard D E, DeBoer J, Stump G S, Ho A D and Seaton D T 2013 Studying learning in the worldwide classroom: Research into edX's first MOOC *Res. Pract. Assess.* **8** 13–25
[23]  Alexandron G, Ruiperez-Valiente J A, Chen Z, Pedro J. Muñoz-Merino and Pritchard D E 2017 Copying @ Scale: Using Harvesting Accounts for Collecting Correct Answers in a MOOC *Comput. Educ.* **108**
[24]  Kim J, Guo P J, Seaton D T, Mitros P, Gajos K Z and Miller R C 2014 Understanding In-Video Dropouts and Interaction Peaks in Online Lecture Videos *Learn. Scale, 2014* 31–40
[25]  Chen Z, Lee S and Garrido G 2018 Re-designing the Structure of Online Courses to Empower Educational Data Mining *Proceedings of 11th International Educational Data Mining Conference* ed K Elizabeth Boyer and M Yudelson (Buffalo, NY) pp 390–6
[26]  Lin S Y, Aiken J M, Seaton D T, Douglas S S, Greco E F, Thoms B D and Schatz M F 2017 Exploring physics students' engagement with online instructional videos in an introductory mechanics course *Phys. Rev. Phys. Educ. Res.* **13** 1–18
[27]  Seaton D T, Kortemeyer G, Bergner Y, Rayyan S and Pritchard D E 2014 Analyzing the impact of course structure on electronic textbook use in blended introductory physics courses *Am. J. Phys.* **82** 1186–97
[28]  Warnakulasooriya R, Palazzo D J and Pritchard D E 2007 Time to completion of web-based physics problems with tutoring. *J. Exp. Anal. Behav.* **88** 103–13
[29]  Rayyan S, Fredericks C, Colvin K F, Liu A, Teodorescu R, Barrantes A, Pawl A, Seaton D T and Pritchard D E 2016 A MOOC based on blended pedagogy *J. Comput. Assist. Learn.* **32** 190–201
[30]  Toven-Lindsey B, Rhoads R A and Lozano J B 2015 Virtually unlimited classrooms: Pedagogical practices in massive open online courses *Internet High. Educ.* **24** 1–12
[31]  Ho A D, Reich J, Nesterko S O, Seaton D T, Mullaney T, Waldo J and Chuang I 2014 HarvardX and MITx: The First Year of Open Online Courses, Fall 2012-Summer 2013 *SSRN Electron. J.* 1–33
[32]  Chen Z, Chudzicki C, Palumbo D, Alexandron G, Choi Y-J, Zhou Q and Pritchard D E 2016 Researching for better instructional methods using AB experiments in MOOCs:Results and Challenges *Res. Pract. Technol. Enhanc. Learn.* **11**
[33]  Mikula B D and Heckler A F 2017 Framework and implementation for improving physics essential skills via computer-based practice: Vector math *Phys. Rev. Phys. Educ. Res.* **13** 010122
[34]  Heckler A F and Mikula B D 2016 Factors affecting learning of vector math from

computer-based practice: Feedback complexity and prior knowledge *Phys. Rev. Phys. Educ. Res.* **12** 010134

[35]  Schroeder N, Gladding G, Gutmann B and Stelzer T 2015 Narrated animated solution videos in a mastery setting *Phys. Rev. Spec. Top. - Phys. Educ. Res.* **11** 010103

[36]  Gutmann B, Gladding G E, Lundsgaard M and Stelzer T Mastery-style homework exercises in introductory physics courses: Implementation matters *Phys. Rev. Phys. Educ. Res.*

[37]  Shi Xia L, Xing Hua L, Hui S and Honesty Cheng Y 2008 METHOD, INTERACTION METHOD AND APPARATUS FOR VISUALIZING HERARCHY DATA WITH ANGULAR CHART

[38]  Ericsson K A, Krampe R T, Tesch-romer C, Ashworth C, Carey G, Grassia J, Hastie R, Heizmann S, Kellogg R, Levin R, Lewis C, Oliver W, Poison P, Rehder R, Schlesinger K and Schneider V 1993 The Role of Deliberate Practice in the Acquisition of Expert Performance *Psychol. Rev.* **100** 363–406

[39]  Ericsson K A, Nandagopal K and Roring R W 2009 Toward a science of exceptional achievement: attaining superior performance through deliberate practice. *Ann. N. Y. Acad. Sci.* **1172** 199–217

[40]  Singh C and Rosengrant D 2003 Multiple-choice test of energy and momentum concepts *Am. J. Phys.* **71** 607

[41]  Thompson K and Yonekura F 2005 Practical guidelines for learning object granularity from one higher education setting *Interdiscip. J. Knowl. Learn. Objects* **1** 163–179

[42]  Bishop C, Yonekura F and Moskal P 2013 Pilot Study Examining Student Learning Gains Using Online Information Literacy Modules *Proceedings of the Association of College and Research Libraries (ACRL) Annual Conference* (Indianapolis, Indiatna) pp 466–71

[43]  Chen Z, Garrido G, Berry Z, Turgeon I and Yonekura F 2018 Designing online learning modules to conduct pre- and post-testing at high frequency *2017 Physics Education Research Conference Proceedings* (Cincinnati, OH: American Association of Physics Teachers) pp 84–7

[44]  Center for Distributed Learning University of Central Florida Obojobo Next

[45]  Instructure Inc. Canvas Learning Management System

[46]  R Core Team 2017 R: A Language and Environment for Statistical Computing *Doc. Free. available internet http//www. r-project. org*

[47]  Wickham H 2017 tidyverse: Easily Install and Load the "Tidyverse"

[48]  Benaglia T, Chauveau D, Hunter D and Young D 2009 mixtools: An R Package for Analyzing Finite Mixture Models *J. Stat. Softw.* **32** 1–29

[49]  Bostock M, Ogievetsky V and Heer J 2011 D3data-driven documents *IEEE Trans. Vis. Comput. Graph.* **17** 2301–9

[50]  Palazzo D J, Lee Y-J J, Warnakulasooriya R and Pritchard D E 2010 Patterns, correlates, and reduction of homework copying *Phys. Rev. Spec. Top. - Phys. Educ. Res.* **6** 010104

[51]  Whitcomb K M, Chen Z and Singh C 2018 Measuring the effectiveness of online problem-solving tutorials by multi-level knowledge transfer *Proc. 2018 Phys. Educ. Res. Conf.* 1–4

# Appendix I: Sample problems used in the AC of each module.

The AC of each module contains a set of isomorphic problems. Only the first problem of the set is shown.

## Module 1:

### Problem 1

A 4 kg object is moving at a constant velocity of 20 m/s in the $+x$ direction. Which of the following objects have close to $\frac{1}{4}$ the kinetic energy of this given object?

Select all the correct statements

Pick **all** of the correct answers

- ☐ A 1 kg object moving at velocity 20 m/s in the $+x$
- ☐ A 4 kg object moving at 6 m/s in the $-x$ direction, and at 8 m/s in the $+y$ direction
- ☐ A 4kg object moving at 5 m/s in the $+y$ direction.
- ☐ A 2 kg object moving at velocity of 10 m/s in the $+x$ direction
- ☐ A 4 kg object moving at 10 m/s in the $-y$ direction

### Problem 2

Which of the scenarios below has the **most** amount of kinetic energy?

A. A 3 kg object moves at an instant velocity of 5 m/s in the $+y$ direction with an acceleration of $6m/s^2$, in the $-y$ direction

B. A 2 kg object moves with an instant velocity of 6 m/s and is also being pushed by a force of 100 N, both in the $+x$ direction

C. A 3 kg object moves with an instant velocity of 4 m/s with an acceleration of 1 m/s both in the $-y$ direction

Pick the correct answer

- ○ All scenarios have the same amount of kinetic energy
- ○ Scenario C
- ○ Scenario B
- ○ Scenario A

## Module 2:

### Problem 1

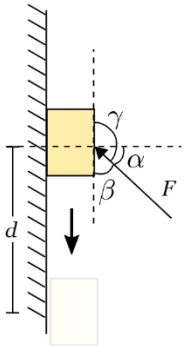

A block is being pressed against a vertical wall with external force $\vec{F}$ and sliding down (as shown in the figure). When it slides down for a distance of $d$, what is the expression for the work done by the external force $\vec{F}$ during the process?

Pick the correct answer

- ○ $Fd\sin(\gamma)$
- ○ $Fd\cos(\gamma)$
- ○ $Fd\cos(\beta)$
- ○ $Fd\cos(\alpha)$
- ○ $Fd\sin(\alpha)$

### Problem 2

The work done by the external force on the block is:

Pick the correct answer

- ○ Negative
- ○ Cannot determine
- ○ Zero
- ○ Positive

## Module 3:

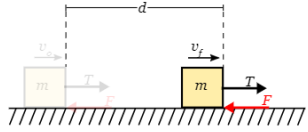

A 6 kg box is being pulled across a rough horizontal floor by a rope with constant tension of T = 8N. A frictional force is opposing the movement with $F_f$ = 6N. The box is being pulled for 4 m and as a result its speed increased to 1.8 m/s. Find the initial velocity during this period.

Pick the correct answer

- ○ 0.75 m/s
- ○ None of these are correct.
- ○ -0.53 m/s
- ○ 0.53 m/s
- ○ 1.60 m/s

## Module 4:

### Problem 1

In each of the following five situations, a small object starts from a certain height and ends up on level ground. When comparing the changes in gravitational potential energy of the object, which of the following statements are true?

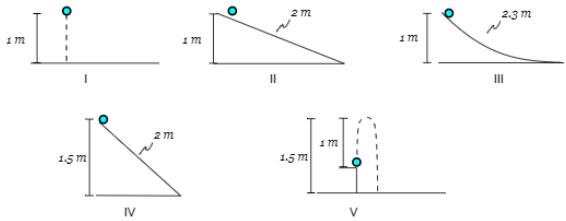

A small object rolls down from different ramps

Pick **all** of the correct answers

- ☐ III>II>I
- ☐ IV=V>I
- ☐ I=II=III
- ☐ IV>I>V
- ☐ I need to study this one
- ☐ III>II=IV
- ☐ None of the above

### Problem 2

A block sitting on a frictionless surface is attached to a spring of spring constant $k$. The other end of the spring is attached to the wall. The equilibrium length of the spring is $x_1$. The spring is first compressed to a distance of $x_2$, then released and extended to a length of $x_3$. The elastic potential energy stored in the spring when it is extended to $x_3$ can be calculated as:

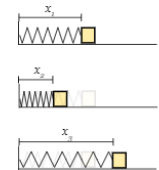

A small object comes down from a certain height

Pick the correct answer

- ○ None of the above
- ○ I need to study for this one.
- ○ $PE = \frac{1}{2}k(x_3 - x_1)^2$
- ○ $PE = \frac{1}{2}k(x_3 - x_2)^2$
- ○ $PE = \frac{1}{2}k(x_3^2 - x_2^2)$
- ○ $PE = \frac{1}{2}k(x_3^2 - x_1^2)$
- ○ $PE = \frac{1}{2}kx_3^2$

## Module 5:

### Is Mechanical Energy Conserved?

Select all the correct statements about the mechanical energy of the cases below (select all that are true).

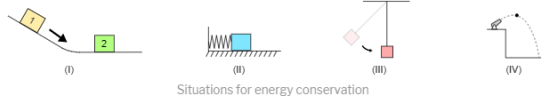

Situations for energy conservation

Case (I) : A block slides down a frictionless ramp, hits a second block on a frictionless surface, and sticks to it.

Case (II) : A block sitting on a **rough** surface is compressed against a spring and then released.

Case (III): A block is tied to a massless string and swings down from a certain height.

Case (IV): A cannonball is launched by a cannon on the edge of a cliff, air resistance is negligible.

Pick **all** of the correct answers

- ☐ Let me take a look at the instructions.
- ☐ In Case IV, after the cannonball is launched and before the cannonball hits the ground, the ME of the system is **not** conserved because there is an external force from the cannon on the ball that makes it fly forward.
- ☐ In Case I the ME of the two blocks is conserved because all the surfaces are frictionless and there are no external forces on either block.
- ☐ In Case II the ME of the spring and the block system is not conserved because the surface is rough.
- ☐ In Case III the ME of the block is not conserved because tension from the spring is a non-conservative external force.
- ☐ None of the statements are correct.

## Module 6:

### Problem 1
#### Stone from cliff

Two identical stones are shot from a cliff from the same height and with identical initial speeds $v_0$. The stone on the left is shot vertically up, and the stone on the right is shot vertically down (see figure). Which one of the following statements best describes which stone has a larger speed right before it hits the ground?

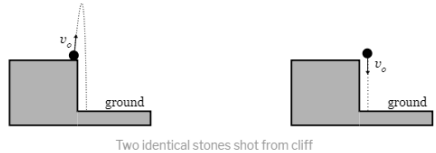

Two identical stones shot from cliff

**Pick the correct answer**

- ○ The stone on the right, because it travels a shorter distance and so is not subjected to negative acceleration.
- ○ None of the statements are correct.
- ○ The stone on the left, because it travelled a longer downward distance so it has more time to accelerate.
- ○ The stone on the left, because it takes a longer time to reach the ground and so has more opportunity to accelerate.
- ○ Both stones have the same speed.

### Problem 2
#### Trapeze Artist

Two trapeze artists are launched out of a cannon with the same initial speed. Both are aimed at a bar that spans the length of the ceiling. Artist A weighs 50 kg, while Artist B weighs 60 kg. Looking at the diagram below, which artist will have a greater speed when they reach the target bar?

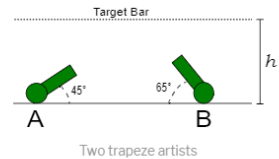

Two trapeze artists

**Pick the correct answer**

- ○ Artist B
- ○ Let me take a look at the instructions.
- ○ Artist A
- ○ Not enough information to decide.
- ○ Both artists will have the same amount of speed when they reach the target bar.

# Module 7

## Problem 1

A block of mass m is compressed against a spring of spring constant $k$. The block is then released and travels up a frictionless incline reaching a maximum height of $h$. The maximum compression of the spring before the block was released was $\Delta x$. The velocity of the block as it travels across the horizontal surface is $v$.

Which of the following statements are true?

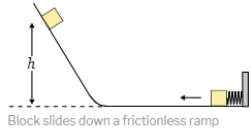

Block slides down a frictionless ramp

**Pick all of the correct answers**

- [ ] If the block started with a compression of $\frac{1}{2}\Delta x$, the block's velocity on the horizontal surface would be $\frac{1}{4}v$.

- [ ] When the block slides back down from the ramp, it will compress the spring again by a maximum amount of $\Delta x$.

- [ ] If the block's velocity on the horizontal surface was $\frac{1}{4}v$, it would reach a maximum height of $\frac{1}{2}h$.

- [ ] If the block were to start with a compression of $2\Delta x$, the block would reach a maximum height of $2h$.

- [ ] Let me take a look at the instructions.

- [ ] None of the above is correct

## Problem 2

A block of mass $m = 0.4kg$ starts from rest and slides down a frictionless ramp, before hitting a spring of spring constant $k = 300N/m$ placed on a horizontal surface. When the block is at the height of $1.5m$ above the ground, its velocity is $2m/s$. What is the maximum compression of the spring? Take $g = 9.8m/s^2$.

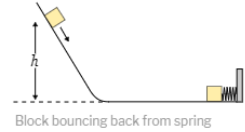

Block bouncing back from spring

**Pick the correct answer**

- ○ Let me take a look at the instructions.
- ○ 0.042 m
- ○ 0.205 m
- ○ 0.211 m
- ○ 0.045 m
- ○ None of these are correct

# Module 8

A small block of mass $m = 0.2kg$ hangs under an elastic rubber band that can be treated as an ideal spring with spring constant $k = 20N/m$. The block and the band were initially held at an angle of $\theta = 60°$ with respect to the vertical direction, and the rubber band was initially held at its equilibrium distance $s_0 = 0.1m$. The block is then released and swings downward. When the rubber band is in the vertical direction, it is measured that the band is extended to $s = 0.25m$. What is the magnitude of the block's velocity when the rubber band is in the vertical position? (use $g = 10m/s^2$ in the calculation)

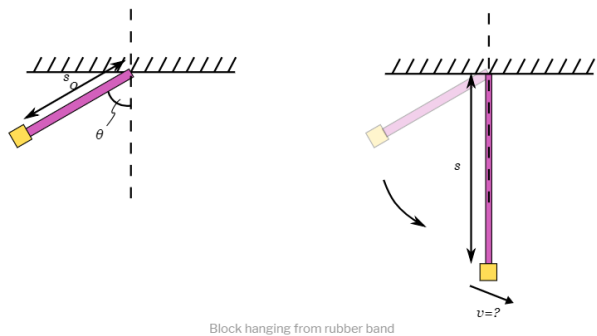

Block hanging from rubber band

**Pick the correct answer**

- ○ None of these are correct.
- ○ $\sqrt{2.25}$m/s
- ○ 2m/s
- ○ $\sqrt{0.75}$m/s
- ○ Let me take a look at the instructions.
- ○ $\sqrt{1.75}$m/s

# Module 9

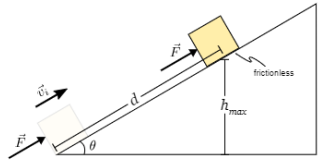

A block of mass $m$ is pushed up a frictionless ramp with initial velocity $v_i$ and constant force $F$. The surface of the wedge forms an angle of $\theta$ with respect to the horizontal direction. What is the final velocity after a certain distance $d$?

Pick the correct answer

- $\sqrt{\dfrac{2Fd - 2mgd\sin(\theta) + mv_i^2}{m}}$

- $\sqrt{\dfrac{Fd + mgd\sin(\theta) + mv_i^2}{m}}$

- $\sqrt{\dfrac{2Fd - 2mgd\sin(\theta)}{m}}$

- $\sqrt{\dfrac{2Fd + 2mgd\cos(\theta)}{m}}$

- $\sqrt{\dfrac{2Fd + 2mgd\cos(\theta) + mv_i^2}{m}}$

# Module 10

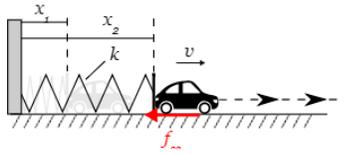

A small toy car of mass $m = 0.05$kg is placed on a rough surface and compressed against a spring with spring constant $k = 20$N/m. The equilibrium length of the spring is $x_0 = 0.25$m, and it is compressed to a length of $x_1 = 0.1$m. After the car is released, the magnitude of kinetic friction between the car and the spring is $f = 1$N. When the spring extends back to a length of $x_2 = 0.2$m, what is the magnitude of the toy car's velocity?

Pick the correct answer

- $v = 2.24$m/s
- $v = 0$m/s
- $v = 3.46$m/s
- $v = 2.8$m/s
- $v = 2$m/s

# Appendix II: Additional Data Analysis Results

1. Duration of students' 1st attempt on each module plotted on log scale. The vertical line is drawn at $t = 40s$.

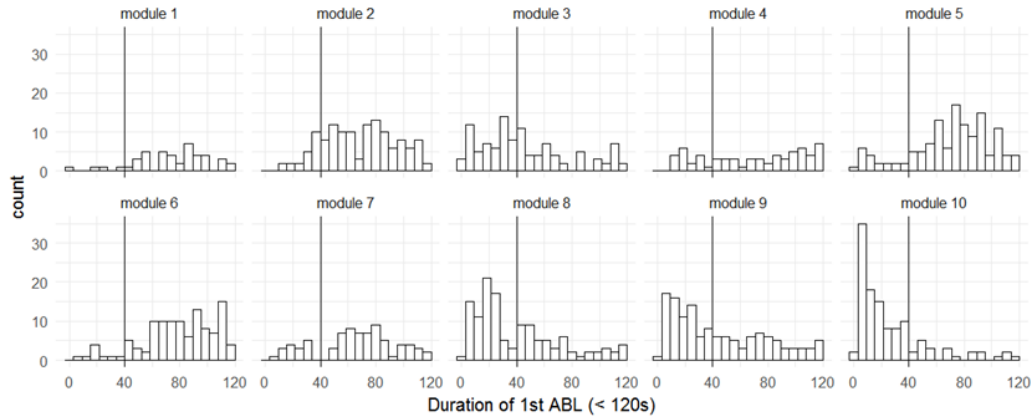

2. Fisher's exact test results for the difference in passing percentage between post-learning attempts > 180s and attempts < 180s on modules 7 – 10.

| Module | p-value | Odds ratio |
|---|---|---|
| 7 | 0.00** | 21.12 |
| 8 | 0.02* | 3.10 |
| 9 | 0.17 | 4.53 |
| 10 | 0.00** | 9.16 |

3. Results of Shapiro-Wilk normality test on MLS log-duration for the 10 modules.

|   | module 1 | module 2 | module 3 | module 4 | module 5 | module 6 | module 7 | module 8 | module 9 | module 10 |
|---|---|---|---|---|---|---|---|---|---|---|
| p | 0.264 | 0.055 | 0.001** | 0.067 | 0.969 | 0.816 | 0.004** | 0.002** | 0.025* | 0.003** |
| W | 0.988 | 0.984 | 0.951 | 0.983 | 0.997 | 0.993 | 0.973 | 0.948 | 0.967 | 0.953 |

4. Time cut-offs between Brief learners and Normal learners for each module.

|   | module 1 | module 2 | module 3 | module 4 | module 5 | module 6 | module 7 | module 8 | module 9 | module 10 |
|---|---|---|---|---|---|---|---|---|---|---|
| cutoff(s) | 219 | 120 | 243 | 156 | 173 | 130 | 294 | 264 | 113 | 101 |

5. Fisher's exact test results for the correlation between Brief/Normal problem solvers on ABL and Brief/Normal learners during MLS. Note that for Odds Ratio cannot be computed for Fisher's exact test for tables larger than 2 by 2 (i.e. Fisher-Freeman-Halton tests).

| Fisher's test | module 1 | module 2 | module 3 | module 4 | module 5 | module 6 | module 7 | module 8 | module 9 | module 10 |
|---|---|---|---|---|---|---|---|---|---|---|
| p | 0.31 | 0.24 | 0.23 | 1.00 | 0.48 | 1.00 | 0.10 | 0.46 | 0.29 | 0.00* |
| odds ratio | 5.04 | 2.3 | 1.93 | 0.97 | 0.31 | 0.64 | 2.61 | 0.6 | 1.87 | 3.46 |

6. Number of students in each engagement mode.

| Engagement mode | module1 | module 2 | module 3 | module 4 | module 5 | module 6 | module 7 | module 8 | module 9 | module 10 |
|---|---|---|---|---|---|---|---|---|---|---|
| NN | 128 | 133 | 70 | 110 | 138 | 88 | 126 | 41 | 57 | 45 |
| NB | 25 | 23 | 12 | 31 | 32 | 23 | 36 | 32 | 13 | 24 |
| BN | 1 | 10 | 18 | 11 | 14 | 6 | 8 | 15 | 21 | 15 |
| BB | 1 | 4 | 6 | 3 | 1 | 1 | 6 | 7 | 9 | 28 |

7. Discussion on Alternative criteria for "Brief" and "Normal" learner categorization.

For the criteria used in the paper, the cutoff between "Brief" and "Normal" learners according to students' MLS duration is determined by the following rule:

If the log of the distribution is best fit with a single normal distribution, than the cutoff is set as one standard deviation below the population mean. If the log of distribution is best fit with multiple normal distributions, the cutoff is set at the intersection between the shortest and the second shortest distribution.

An artifact of using this rule is that the cutoff for modules with multiple normal distributions often tends to be at a longer duration than modules with single normal distributions. Therefore we tested most of our major results using the following two alternative criteria which each resulted in more uniform cutoff times across all 10 modules:

- Alt-cut1: For modules with multi-component log-normal distributions, the cutoff is changed to the mean of the shortest normal component. (i.e. the criteria for "Brief" is stricter)
- Alt-cut2: For modules with single-component log-normal distributions, the cutoff is changed to 0.5 standard deviation below the mean. (i.e. the criteria for "Brief" is more generous)

For each alternative criteria, we tested the correlation between pre-learning attempts and MLS duration, as well as the correlation between four engagement modes and post-learning passing rate. The results are largely identical, except: For Alt-cut1, correlation between Brief pre-learning attempt and Brief learning is statistically significant at $\alpha < 0.01$ level for module 7, but the numer of the Brief pre-learning attempts was very small. For Alt-cut 2, correlation between the four engagement modes and post-learning passing is no-longer significant for module 4 at $\alpha = 0.05$ level, which is reasonable since Alt-cut2 includes more students in the "Brief" learning group, and could have included some stronger students who were seriously learning the material. However, the correlation is significant at the $\alpha = 0.01$ level for module 2.